 \documentstyle[12pt,aasms4]{article}

\newcommand{\kms}{km\thinspace s$^{-1}$}
\newcommand{\Kkms}{K\thinspace km\thinspace s$^{-1}$}

\received{}
\accepted{}
\journalid{}{}
\articleid{}{}
\slugcomment{}

\begin{document}
\title{High Resolution CO Imaging of the Molecular Disk around the  
Jets in  KjPn~8}

\author{T. Forveille}
\affil{Observatoire de Grenoble, B.P. 53X, 38041 Grenoble Cedex, France}
\author{P.J. Huggins}
\affil{Physics Department, New York University, 4 Washington Place,
New York, NY 10003}
\author{R. Bachiller}
\affil{IGN Observatorio Astron\'omico Nacional, Apt 1143, 28800 Alcal\'{a} 
de Henares, Spain}
\and
\author{P. Cox}
\affil{Institut d'Astrophysique Spatiale, Universit\'e de Paris XI,
91405 Orsay, France}
\affil{Institut d'Astrophysique de Paris, 92 bis, Boulevard Arago, 
75014 Paris, France}

\begin{abstract}
We report high resolution (2.5\arcsec) CO $J$=1$-$0 line
imaging which confirms the presence of a molecular disk around the origin 
of the
spectacular, $14\arcmin\times 4\arcmin$, episodic jets in the
planetary nebula KjPn~8.  The disk is 30\arcsec\ in diameter with an
expansion velocity of $\approx 7$~\kms.  The axis of the disk is
aligned with the youngest and fastest ($\approx 300$~ \kms) of the
bipolar jets, and there is evidence for interaction between the jets
and the disk material.  The inner 4\arcsec\ of the disk are
photo-ionized by the central star. The disk-jet system dominates the
environment of this young nebula, and should govern the morphology of
KjPn~8 as it evolves to become fully ionized.

\end{abstract}


\section{Introduction}

The presence of jets in planetary nebulae (PNe) has only recently been
widely recognized, but they prove to be of great interest both for the
study of jet phenomena and for their effects on PN structure.  In PNe
with jets, the action of the collimated flows will modify, and in some
cases may completely change the traditional picture of PN formation by
the two wind model, in which the shaping is produced by the action of
an isotropic wind (e.g., \markcite{ka85}Kahn \& West 1985,
\markcite{ba87}Balick et al.\ 1987).

Evidence for jets in PNe is found in objects at all stages of
evolution.  It includes direct observation of elongated, high velocity
structures, which may be episodic and exhibit changes in direction
(\markcite{lo97}L\'opez 1997), as well as other features in the
nebulae such as point symmetries (\markcite{sa92}Schwarz et al.\ 1992)
and poly-polar structures (\markcite{ma96}Manchado et al.\ 1996),
which are likely produced by jets or related processes.  Observations
suggest that jet activity begins at the earliest stages of PN
formation, where it will have most impact on the structure of the
neutral circumstellar envelope that evolves into the ionized nebula
(\markcite{ha96}Huggins et al.\ 1996).  Direct evidence for
interaction of the jets with the envelope is seen in high velocity
molecular outflows in several proto-PNe and young PNe, the best
studied case being CRL 618 (\markcite{ca89}Cernicharo et al. 1989,
\markcite{na92}Neri et al. 1992, \markcite{ha96}Hajian et
al. 1996). The extent of the jets in these young objects is, however,
usually quite small and not well resolved, so the recent discovery of
very extended jets in KjPn~8 provides a unique opportunity to study
their effects.

KjPn~8 (PN G112.5$-$00.1, K3$-$89) was first identified as a PN by
\markcite{ka71}Kazarian \& Parsamian (1971) and independently by
\markcite{ko72}Kohoutek (1972).  The ionized nebula is small, with an
angular diameter of $\approx 4\arcsec$, but surrounding this
\markcite{la95}L\'opez et al.\ (1995) have discovered an enormous
($\approx 14\arcmin \times 4\arcmin$), low surface-brightness, bipolar
structure.  It consists of symmetric jet features with paired emission
regions, characteristic of episodic ejection (see Fig.\ 1).  Optical
line ratios of the emission indicate shock excitation, and the high
velocities of the jets ($\gtrsim 200$~\kms) have been directly
confirmed by radial velocity measurements (\markcite{la97}L\'opez et
al.\ 1997). Proper motions of the jets have also been detected by
\markcite{me97}Meaburn (1997), and lead to a distance estimate for
KjPn~8 of 1.6 kpc. At this distance, the extended bipolar structure is
$6.5 \times 1.9$~pc in size.

We have recently reported the detection of millimeter CO emission
towards KjPn~8, which reveals the presence of molecular gas near the
origin of the jets (\markcite{ha97}Huggins et al.\ 1997). The
distribution of the gas was only partially resolved by our 12\arcsec\
observations, but suggested a disk morphology.  In view of the
potential importance of mutual disk-jet interactions for understanding
PN formation, we have re-observed the envelope at much higher angular
resolution in CO, and in this {\it Letter} we report the results.

\section{Observations}

The observations were made in the 115~GHz (2.6~mm) CO $J=1-0$ line
during April and May 1997 using the IRAM interferometer (Guilloteau et
al.\ 1992) located at Plateau de Bure, France.  The array consisted of
five 15~m antennas, equipped with SIS heterodyne receivers. The
observations, centered on KjPn~8, were made with two configurations of
the array, C2 and D.  At the frequency of the CO line the primary beam
is 44\arcsec\ (FWHM) and the effective velocity resolution of the line
observations was 1.6~\kms.  Observations of the continuum at 115~GHz
were also made, with a bandwidth of 320~MHz.

The RF passband and amplitude were calibrated using NRAO~530 (flux
5.75~Jy) and 3C454.3 (6.4~Jy), and phase calibration was performed
every 20 minutes using J0102+5824 (2.6 Jy). Data obtained using the
IRAM 30~m telescope were used for the zero spacing observations. The
uv data were Fourier transformed and CLEANed, using the Clark
algorithm, and the restored Gaussian clean beam is $2.8\arcsec \times
2.1\arcsec$ at a position angle of 70\arcdeg.

The results of the CO observations are shown in Figs.\ 2--4.  The
continuum at 115~GHz was not detected with an upper limit of $\lesssim
2$~mJy, consistent with the weak continuum emission of 0.8~mJy at
3.5~cm measured by \markcite{lo95} L\'opez et al.\ (1995).

\section{Results and discussion}
 
\subsection {Properties of the disk}

Our high resolution CO observations provide striking evidence for the
presence of a disk at the center of KjPn~8. The overall distribution
of the molecular gas is shown in the velocity integrated CO map in
Fig.~2.  Note that this figure corresponds to the small square at the
center of Fig.~1.  The CO emission shows a roughly elliptical
morphology, with an overall extent of $\approx 30\arcsec\times
14\arcsec$ at a position angle of 36\arcdeg. There is a distinct
$\approx 4\arcsec$ hole at the center which corresponds in size and
position with the small ionized nebula.

The detailed structure and kinematics of the molecular gas can be seen
in the channel maps in Fig.~3.  Inspection of the maps shows that the
main distribution of CO is symmetric in both position and velocity
about a systemic velocity of $V_{\rm o} = -35$~\kms\ (LSR), in close
agreement with the velocity of the the ionized nebula measured using
optical lines ($V_{\rm o} = -34$~\kms , \markcite{la97}L\'opez et al.\
1997).  At this velocity, the CO emission is divided into two separate
parts, with essentially no emission at the map center.  At more
negative (approaching) velocities, the two regions of CO emission join
together towards the north west, and at more positive (receding)
velocities, they join together towards the south east. This kinematic
structure is the signature of an expanding disk, whose near side tilts
away from earth to the north west.  This is fully consistent with the
integrated intensity map in Fig.~2, where the projected disk is seen
as an elliptical structure.  From the
orientation and dimensions of the ellipse in Fig.\ 2, the axis of the
disk lies at a position angle of 126\arcdeg\ (i.e., along the minor
axis of the ellipse) and is tilted away from the line of sight to the
north west by $\approx 65\arcdeg$.

The CO channel map at the systemic velocity of $V_{\rm o} = -35$~\kms\
corresponds to a cross section of the disk in the plane of the sky.
The disk is not flat.  At the inner radius it is $\approx 4\arcsec$
thick, comparable in size to the ionized nebula, but farther out it
thickens to $\approx 20\arcsec$.  Along the disk axis this structure
corresponds to bipolar cavities on each sides of the center, with
roughly conical shapes and opening angles of 60\arcdeg --90\arcdeg.

Additional perspectives on the disk structure are shown in the
velocity strip maps in Fig.~4. The strips through the center along the
principal axes (Figs. 4a and 4b) show the central cavity which
corresponds to the position of the ionized nebula. The strip along the
direction of the disk axis (4b) also clearly shows the tilt of the
disk system; the difference in velocity between the two sides gives an
expansion velocity for the bulk of the gas of $\approx 7$~\kms\
(corrected for inclination), but much higher velocities (up to 28
\kms) are present near the center, where the surfaces of the disk are
seen almost directly along the line of sight.  Irregularities in the
disk are also seen in the data, the most remarkable being hollow
structures within the disk, shown in the strip in Fig.~4c.

The CO emission can also be used to estimate the mass of the disk.
For optically thin, LTE emission, the peak 1$-$0
intensity of 55~\Kkms\ corresponds to a CO column density of
$8 \times 10^{16}$~cm$^{-2}$ for an excitation temperature of 25 K 
which is typical of that found in PNe (\markcite{ba93}Bachiller et al. 
1993). The total flux in the line summed over the disk is
9,600~\Kkms\,arcsec$^2$. Using a representative abundance of ${\rm
CO/H} = 3\times10^{-4}$ and a distance $d = 1.6$~kpc, this corresponds
to a mass of $M_{\rm m} \gtrsim 0.03\ M_{\sun}$, in good agreement
with our previous estimate based on the 2$-$1 line
(\markcite{ha97}Huggins et al.\ 1997).

The observed flux in the 1$-$0 line is, in fact, similar to that in
the 2$-$1 line, which suggests that the CO emission is optically
thick. Thus our estimate for $M_{m}$ is a lower limit. The high
opacity also suggests an explanation for the asymmetry of the disk
seen in Fig.\ 2, in which the receding limb is brighter than the
approaching limb.  If the excitation decreases with distance from the
star, as expected, the hotter, inner surface of the receding limb
will appear brighter than the outer surface of the approaching limb,
as observed.

For comparison with the mass of the disk, the mass of the ionized
nebula estimated from the radio continuum is $M_{i} = 5\times 10^{-4}
\ M_{\sun}$ for $d=1.6$~kpc (\markcite{ha97}Huggins et al.\
1997). Thus the ratio $M_{m}/M_{i}$ is large ($ \gtrsim 60$), and the
mass is dominated by the mass of the molecular disk.

\subsection{Interaction of the disk and the jets} 
 
Our observations of the disk in KjPn~8 provide important information
on the region where the jets originate. Optical images reveal several
different jet axes (see Fig.~1) defined by point symmetric emission
regions that are produced by episodic ejections with the rotation or
wandering of the jets.  The longest jet axis (labeled C by
\markcite{lo95}L\'opez et al.\ 1995) lies at a position angle of
71\arcdeg, and the shortest jet axis, with the most intense emission
(axis A), lies at a position angle of 121\arcdeg.

The orientation of the molecular disk that we measure (Sect.\ 3.1) is
indicated in Fig.~1, and clearly corresponds with jet axis A.  In
addition, according to \markcite{lo97}L\'opez et al. (1997), axis A is
tilted by 60\arcdeg\ away from the line of sight to the north west, in
good agreement with our estimate for the tilt of the molecular disk
axis. We therefore conclude that jet A and the disk axis are aligned
in space, and are physically related. The expansion time scales,
$\tau=R/V$, where $R$ is the characteristic size and $V$ is the
expansion velocity, are consistent with this.  For the A jets, $\tau_j
\approx 2,700$~yr (\markcite{lo97}L\'opez et al.\ 1997); for the
(half-size) of the molecular disk $\tau_d \approx 8,000$~yr; and for
ionized inner disk, $\tau_i \lesssim 2,000$~yr.

In optical images of KjPn~8 (\markcite{lo95}L\'opez et al.\ 1995),
features which connect the heads of the A jets to the center indicate
quite a large opening angle $\approx 45\arcdeg$--60\arcdeg, which fits
in well with the large opening angle of the bipolar cavities formed by
the molecular disk. Interaction with the disk is signaled by the high
velocity molecular gas (the extreme channels in Fig.~3) which is seen
at the rims of the disk, and is likely entrained in the sides of the
jets.  A similar situation probably also exists in the ionized gas
closer ($\lesssim 2\arcsec$) to the star. The bulk of the ionized gas
has a low expansion velocity ($V_{\rm e} = 5$--7.5~\kms,
\markcite{sa86}Sabbadin et al.\ 1986) similar to that of the molecular
gas, and represents the photo-ionized inner disk, but high velocity
50--60~\kms\ components of [N II] 6584\,\AA\ can also be traced 
into the center (\markcite{lo97}L\'opez et al.\ 1997).

Given the thinness of the disk relative to the extent of the jets, it
seems unlikely that the molecular gas is responsible for focusing the
jets, as has been suggested in the case of He 3-1475 by
\markcite{ba97}Borkowski et al.\ (1997).  A more likely scenario is
that a common mechanism induced the mass-loss in the equatorial plane
and created the polar jets close to the center -- probably in a binary
or multiple central star system.  Any neutral gas ejected at high
latitudes will have been evacuated along the polar axis by the action
of the jets, and the lateral pressure in the jet would naturally
expand to fill the opening angles of the disk. Consistent with this,
part of the structure in the disk appears to be compressed edges or
rims. The more irregular structure may result from instabilities, or
from variations in the disk ejection process.

The origin of the wandering or precession of jets in PNe is not clear
(see, e.g.,\markcite{li97} Livio \& Pringle 1997 for a recent model), but our
observations provide useful constraints in the case of KjPn~8.  From
the geometry and kinematics of the disk there is no evidence for
alignment or interaction of the molecular gas with the most extended
jets along axis C. These are the oldest jets, with an expansion time
scale of $\sim 16,000$~yr, consistent with the idea that they predate
the molecular disk. The shape of the envelope of excited gas which
surrounds these jets (see, Fig.\ 2 of \markcite{lo95}L\'opez et al.\
1995) also shows no trace of the molecular disk orientation, and, if
anything, suggests a waist perpendicular to the C axis.  This argues
against a commonly assumed scenario in which the jets rotate or
precess relative to a well defined equatorial plane, and it rather
suggests that the jets and their associated equatorial material have
been ejected in a series of different directions during the evolution
of the system.

Our observations are also relevant to the future evolution of KjPn~8.
The observed structure means that the young ionized nebula at the
center of the system is currently surrounded by a reservoir of neutral
gas in the disk plane and evacuated poles along the axis of the
disk. Whether or not the jets continue to be active, the environment
of the nebula is largely determined by these structural features, and
they are likely to dominate the future morphological evolution of the
nebula.

\section{Conclusions}

Our molecular line imaging reveals in detail the structure and kinematics of 
the
molecular disk around the origin of the remarkable jets in KjPn~8.
The axis of the disk is aligned with the youngest of the bipolar jets
and there is evidence for disk-jet interaction.  The disk and jets are
the major structural features around the young nebula, and are likely
to dominate the morphology as KjPn~8 evolves further into a fully
ionized PN.

\acknowledgements
We are grateful to the IRAM staff for obtaining the observations, and for 
their excellent support during the data reduction.
This work has been supported in part by NSF grant AST-9617941 (to P.J.H.).



\clearpage

\begin{figure}
\caption[]{
Outline of the main optical features of KjPn~8, adapted from the
H$\alpha$ image by \markcite{la95}L\'opez et al.\ (1995).  
The small square at the center
corresponds to the map in Fig.~2. The diagonal lines indicate the
orientation of the molecular disk axis (see text) 

}
\end{figure}

\begin{figure}
\caption[]{
Map of the CO(1$-$0) emission in KjPn~8 in integrated intensity. 
The co-ordinates of the map center are 23:24:10.440 60:57:31.00 (J2000.0).
The cross indicates the center of the ionized nebula, and  
the inset shows the beam size. 
The first solid contour and the contour spacing are 7.5~\Kkms;
the dashed contour is negative with the same spacing
}
\end{figure}

\begin{figure}
\caption[]{
Channel maps in the CO(1$-$0) line. The channel separation is 1.6~\kms. 
The lowest contour
and the contour intervals are 0.5~K
}
\end{figure}

\begin{figure}
\caption[]{
Velocity-strip maps in the CO(1$-$0) line. 
a) Perpendicular to the disk axis, through the center.
b) Along the disk axis. 
c) Parallel to the disk axis through offset 3\arcsec, 5\arcsec.
The lowest contour and the contour intervals are 0.4~K

}
\end{figure}

\onecolumn
\input psfig.sty
\clearpage
\vskip 14.0cm
\centerline{\psfig{file=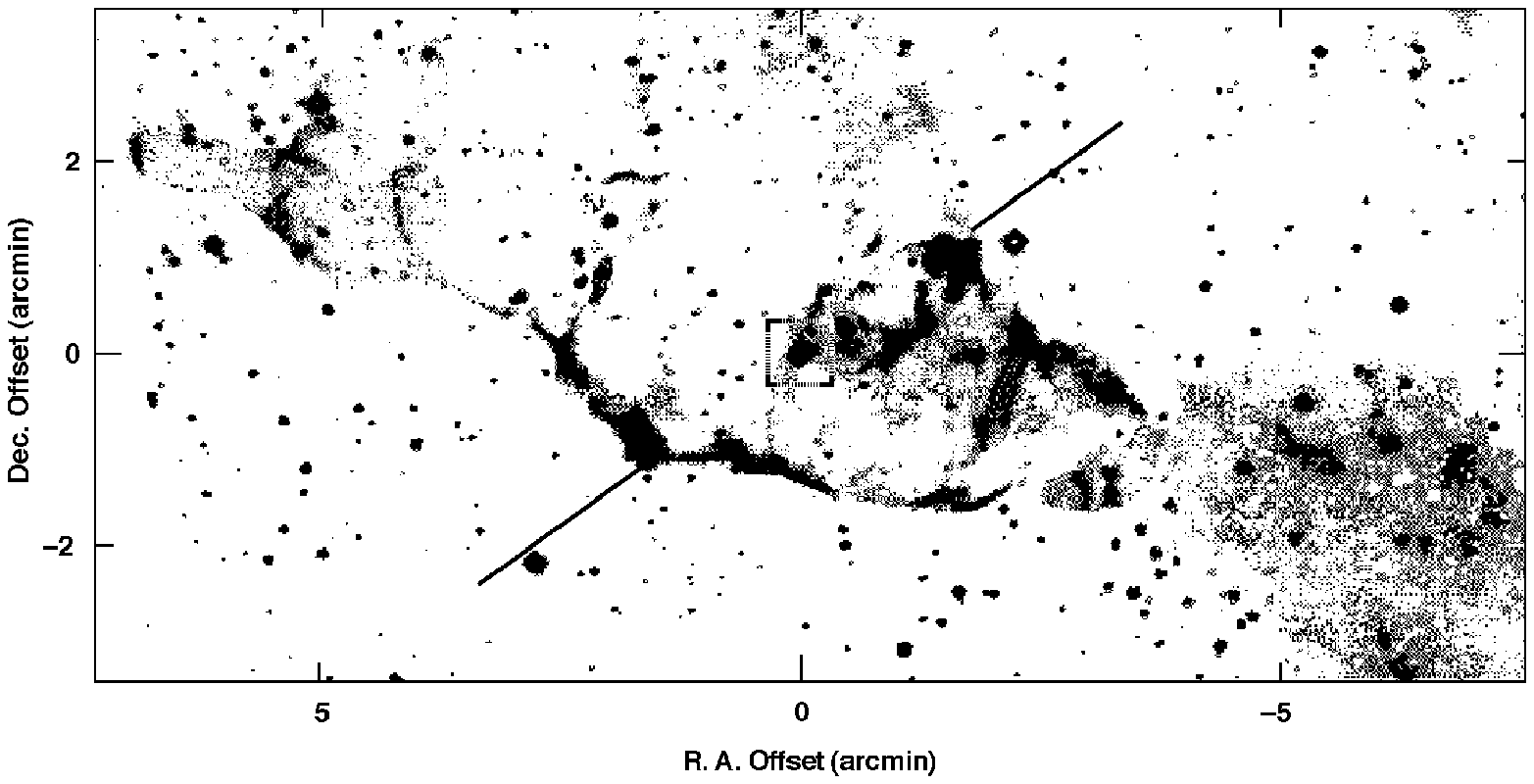,height=17.0cm,angle=0}}
\clearpage
\vskip 14.0cm
\centerline{\psfig{file=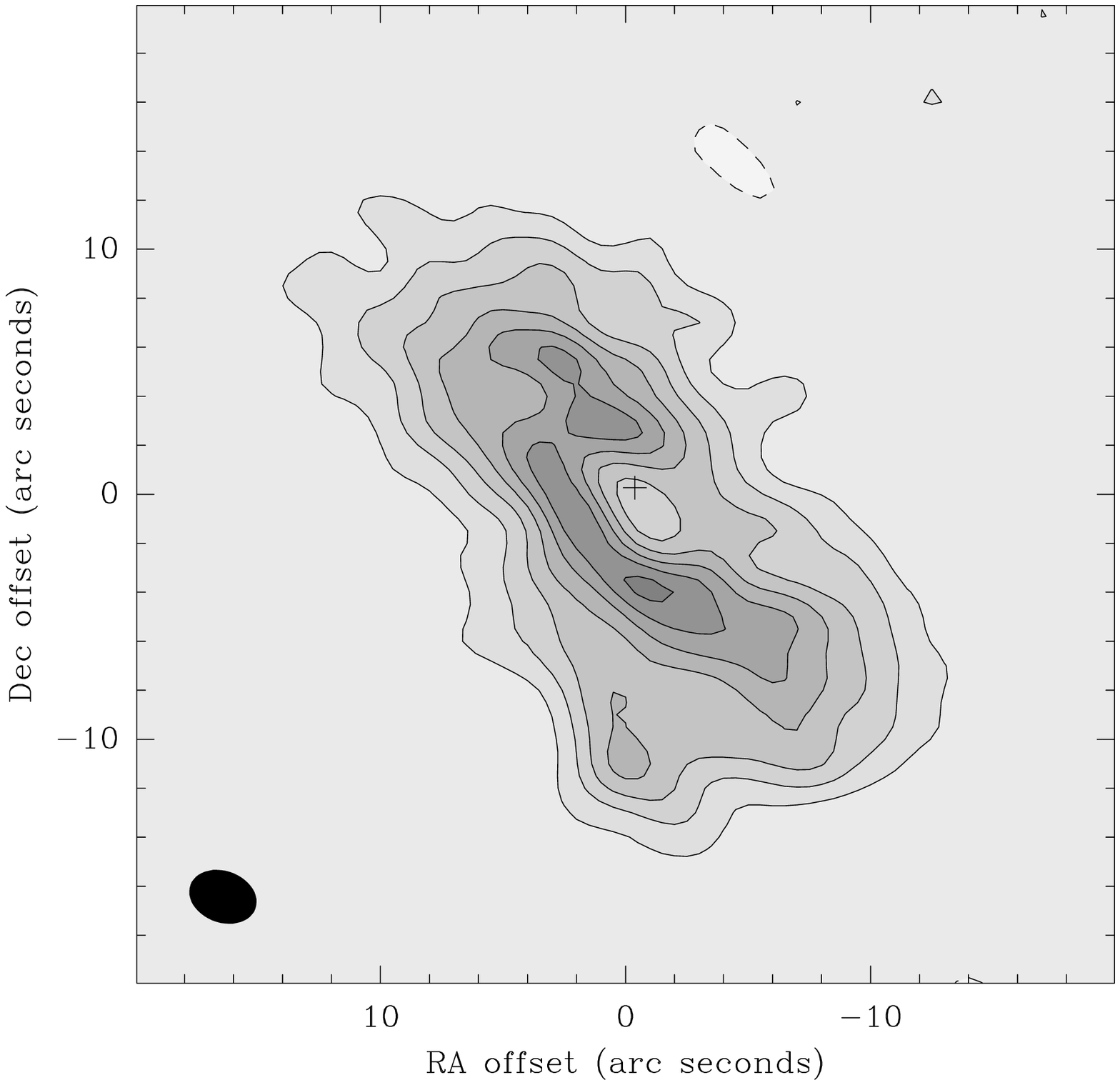,width=18.0cm,angle=270}}
\clearpage
\vskip 14.0cm
\centerline{\psfig{file=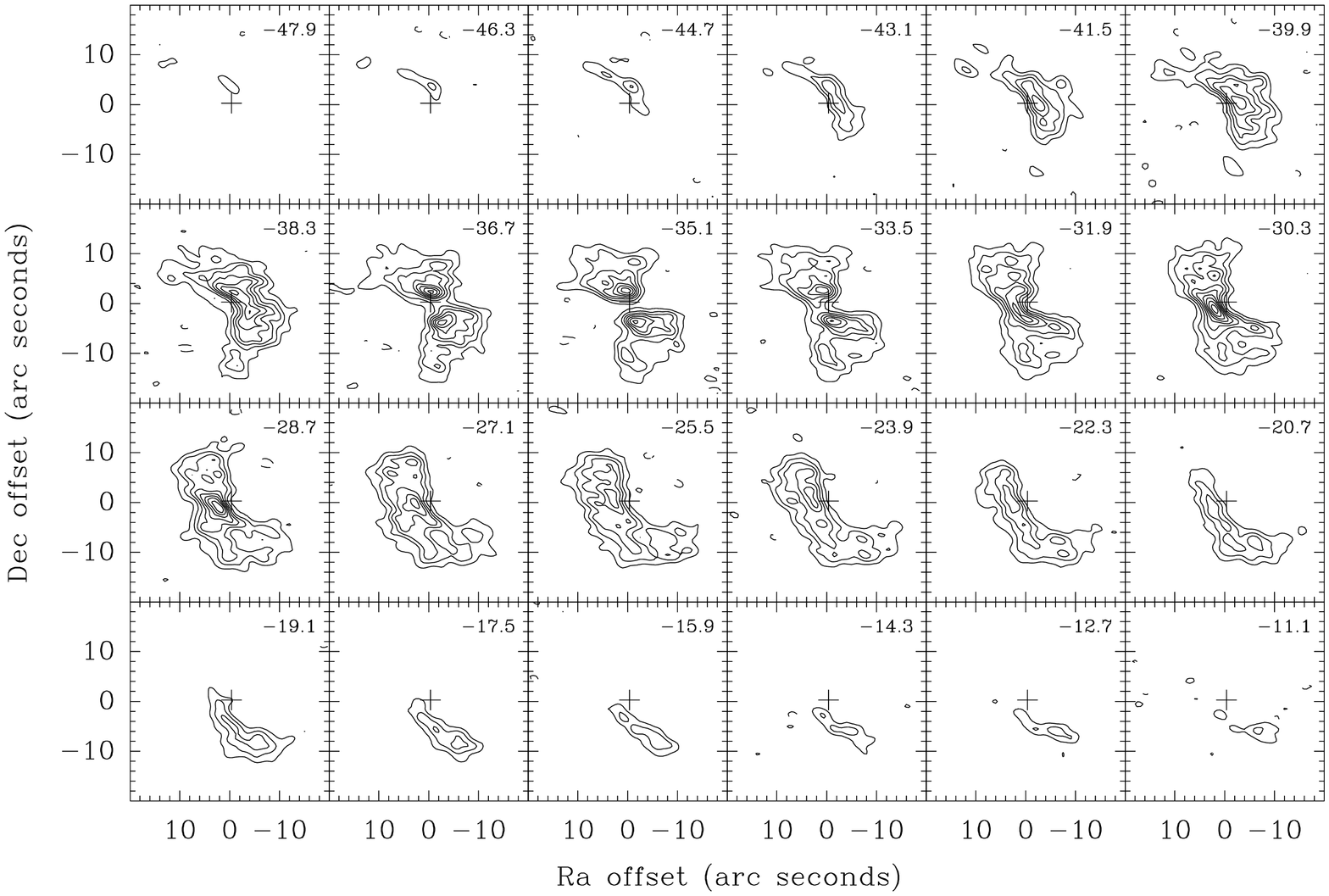,width=18.0cm,angle=270}}
\clearpage
\vskip 14.0cm
\centerline{\psfig{file=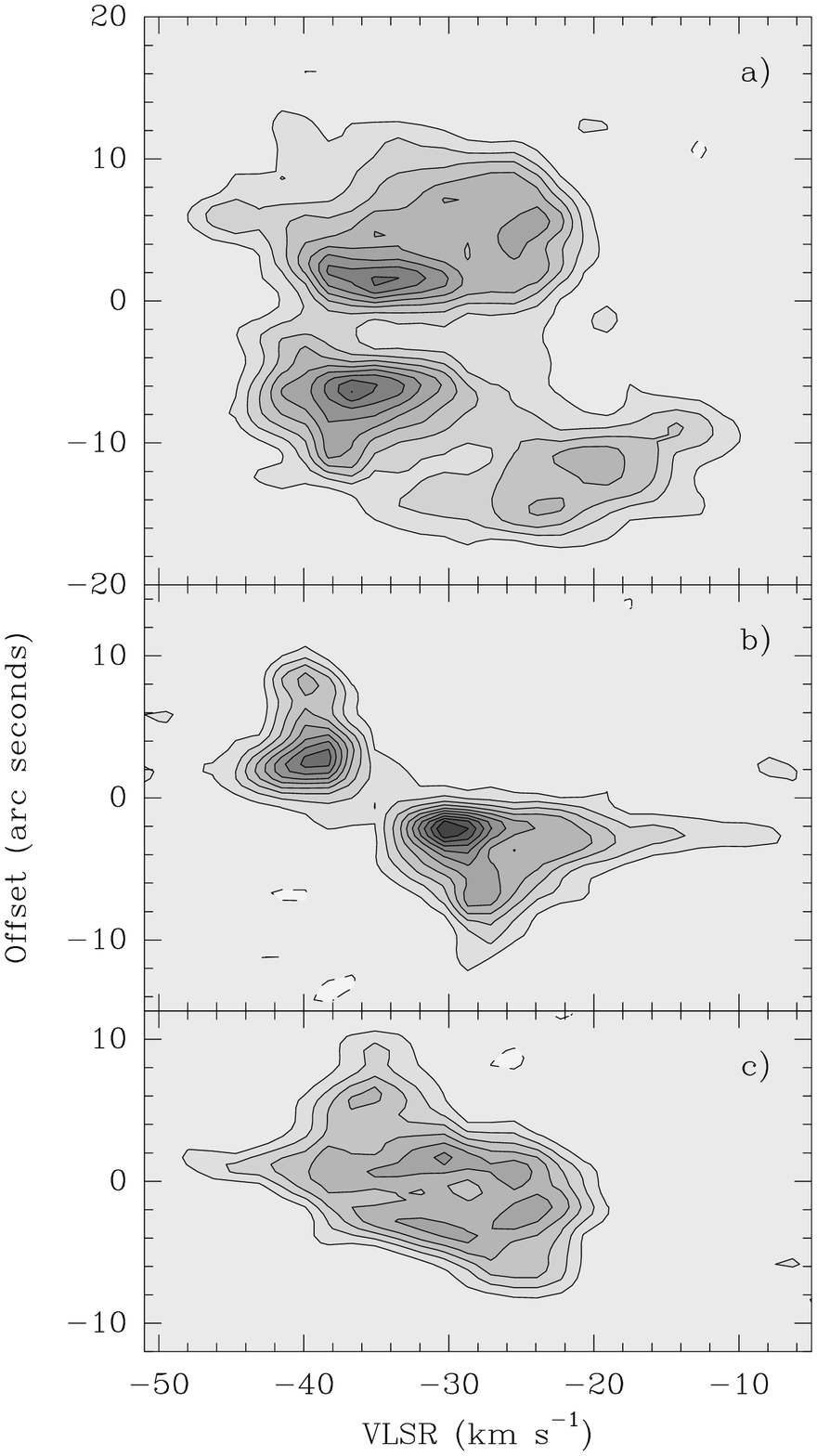,height=24.0cm,angle=0}}
\clearpage

\end{document}